\documentclass[12pt]{iopart}

\usepackage{graphicx}
\usepackage{iopams} 
\begin{document}

\title[Correlation effects in the iron pnictides]
{Correlation effects in the iron pnictides}

\author{Qimiao Si$^{1}$, Elihu Abrahams$^{2}$, Jianhui Dai$^{3}$,
Jian-Xin Zhu$^{4}$}

\address{$^{1}$ Department of Physics and Astronomy, Rice University,
Houston, Texas 77005, USA}
\address{$^{2}$ Center for Materials Theory, Department of Physics and
Astronomy, Rutgers University, Piscataway, New Jersey 08855, USA}
\address{$^{3}$ Zhejiang Institute of Modern Physics,
Zhejiang University, Hangzhou 310027, China}
\address{$^{4}$ Theoretical Division, Los Alamos National Laboratory, 
Los Alamos, New Mexico 87545, USA}
\begin{abstract}
One of the central questions about the iron pnictides 
concerns the extent to which their electrons are strongly
correlated. Here we address this issue through the phenomenology
of the charge transport and dynamics, single-electron excitation
spectrum, and magnetic ordering and dynamics. We outline the 
evidence that the parent compounds, while metallic, have electron
interactions that are sufficiently strong to produce
{\it incipient} 
Mott physics. In other words, in terms of the strength
of electron correlations compared to the kinetic energy, 
the iron pnictides are closer to intermediately-coupled systems
lying at the boundary between itinerancy and localization,
such as ${\rm V_2O_3}$ or Se-doped 
${\rm NiS_2}$,
rather than to simple antiferromagnetic metals like Cr. 
This level of electronic correlations produces a new small
parameter for controlled theoretical analyses, namely 
the fraction of the single-electron spectral weight that
lies in the coherent part of the excitation spectrum. Using
this expansion parameter,
we construct the effective low-energy Hamiltonian
and discuss its implications for the magnetic order
and magnetic quantum criticality. Finally, this approach 
sharpens the notion of magnetic frustration for such 
a metallic system, and brings about a multiband 
matrix $t$-$J_1$-$J_2$ model for the carrier-doped iron pnictides.

\end{abstract}

\maketitle

\section{Introduction}
Two decades after the discovery of the cuprate superconductors, 
we finally have a new 
and copper-free family of 
materials with high temperature 
superconductivity \cite{Kamihara_FeAs,Zhao_Sm1111_CPL08}.
Naturally, these iron pnictides have generated 
an enormous interest, both theoretical and
experimental. One of the intriguing features of the iron pnictides 
is the phase diagram itself. Typically, superconductivity
arises from carrier doping of the parent iron arsenides that
are metallic antiferromagnets. 

A central question is how strongly correlated are the iron pnictides?
It is meaningful to address the issue in the normal state,
given 
that the energy scales for the normal state electronic
and magnetic excitations appear to be considerably
larger than those for superconductivity.
Moreover, it is adequate to focus on the 
3$d$ electrons of the Fe atoms, 
which represent the majority 
of the electronic states near the Fermi energy.
There is an important simplicity here,
namely the Fe atoms, whose 3$d$ orbitals dominate the states near the 
Fermi energy, 
are arranged into square-lattice layers.
But there is also
complication: several if not all
of the five 3$d$ orbitals need to be taken into
account \cite{Lebegue, Singh, Haule, Hirschfeld, Kuroki,Raghu,Lee_Wen}.

\begin{figure}[htbp]
\centering
\includegraphics[width=6.0in]{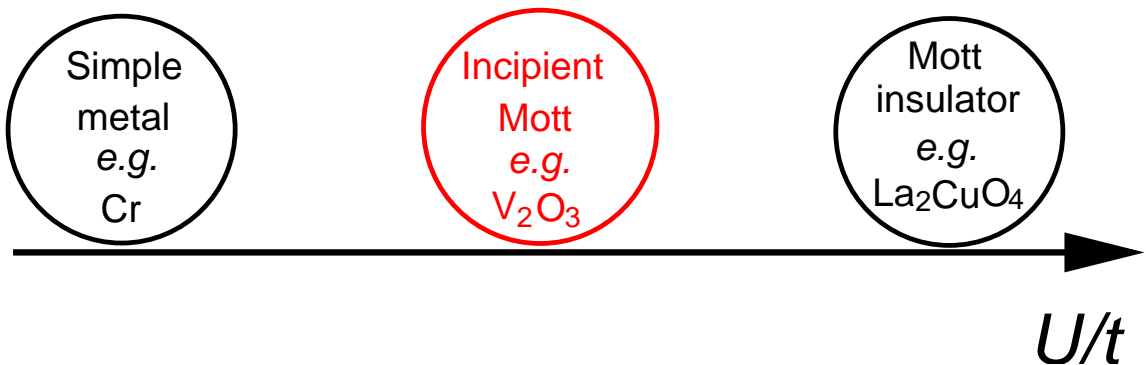}
\caption{Different reference regimes corresponding to different ranges
of interaction strength. When the interaction is large compared to the 
Fermi energy, the system is in a Mott-insulating state. In the opposite
limit, it is a simple metal; Cr is a canonical example for this regime.
For intermediate values of interaction 
close to but below the Mott transition threshold, the system is
metallic but has a sizable fraction of single-electron spectral
weight in an incoherent part as a precursor to the Mott localization;
${\rm V_2O_3}$ and ${\rm NiS_{2-x}Se_x}$ are known systems belonging to 
this regime.
}
   \label{fig:correlation-placement}
\end{figure}

Consider the parent iron pnictides, which may be described in terms
of six electrons occupying the nearly degenerate 3$d$ orbitals at each site
of the square lattice. We can frame the discussion about the strength
of electron correlations in terms of Fig.~\ref{fig:correlation-placement},
which divides partially filled lattice electron systems with integer 
number of electrons into three broad categories:
\begin{itemize}
\item[*] When the interactions are strong compared to the kinetic
energy or bandwidth (large $U/t$ in Fig.\ 1), the system would 
in general be a 
Mott insulator. This would be a muti-orbital analog
of what happens in the undoped cuprates,
and it clearly does not apply here since 
the parent iron arsenides are metallic.
\item[*] When the interactions are relatively weak compared to the kinetic
energy, we have a simple 
metal. This would be the analog of Cr \cite{Fawcett_rmp1,Fawcett_rmp2}.
\item[*] When the interactions are comparable to the kinetic energy, 
the system is metallic but is on the verge of localization. Typical
examples of such incipient Mott systems are ${\rm V_2O_3}$
and Se-doped ${\rm NiS_2}$ \cite{Imada_rmp98}.
\end{itemize}
In the following, we extend our previous 
considerations \cite{Si_prl08,Dai_08}
and provide the phenomenological arguments that
the iron pnictides are very different 
from the weakly-correlated cases like Cr,
but are much more similar to the incipient Mott 
systems such as ${\rm V_2O_3}$
and Se-doped ${\rm NiS_2}$.
We then discuss how this placement of the iron pnictides 
at the boundary between itinerancy and localization serves as the basis
to construct the low-energy effective Hamiltonian.
The latter, in turn, allows a controlled analysis of the magnetic order
and magnetic fluctuations in the iron pnictides.

\section{Simple metals vs. metals proximate to Mott localization}

There are several distinct aspects of physics in the different regimes
discussed above. The degree of itinerancy and incipient localization
is by definition different (see below).
Relatedly, each regime has its own natural small parameter
for controlled theoretical approaches. Finally, the difference in the
small parameter in turn leads to different ways of treating the magnetism.

When the interaction is weak compared to the bandwidth (or Fermi energy),
it is standard to use the non-interacting limit as the reference point,
and adopt the interaction normalized by the kinetic energy as a small 
parameter for perturbation theory. In spite of the weak interaction, 
antiferromagnetism can still develop when the degree of Fermi-surface
nesting is sufficiently large. Cr is the canonical example for this.

When interaction is of the order of the Fermi energy, the interaction 
normalized by the kinetic energy is no longer a small quantity. 
In order to treat the effect of interactions in a controlled fashion,
we need to use a different reference point. One possibility
is the point of the Mott transition, where the coherent electronic 
excitations have
vanished and all the electronic excitations are incoherent.
If we denote by $w$ the fraction of the single-electron spectral weight
that lies in the coherent part,
$w$ then serves as an emergent expansion parameter of the theory.
This quantity provides a precise definition of the degree of itinerancy
introduced earlier: a larger $w$ means a larger degree of itinerancy and,
by extension, a smaller degree of incipient localization.

In this approach, the reference point, at $w=0$, already captures 
the magnetism associated with the localized moments. When $w$ is non-zero,
a perturbative treatment in $w$ will lead to an effective low energy 
theory that couples the local moments
to the itinerant coherent electronic 
excitations. Following our earlier work \cite{Si_prl08,Dai_08}, 
we will give the details of this construction 
in Sec.~\ref{Sec:Mag}.

Physically, the incoherent excitations are non-perturbative effects 
that go beyond the Fermi liquid theory. They represent a precursor 
to the lower and upper Hubbard bands in a Mott insulator, which itself
is a phenomenon non-perturbative in interactions. 
As we shall see, the existence of these incoherent
single-electron excitations will also have important consequences
for the description of magnetic order and magnetic dynamics.

Our placement of the iron pnictides at the boundary
of itinerancy and localization is guided by phenomenology.
Related intermediate-coupling studies have also appeared in the literature
\cite{Haule,Fang:08,Xu:08,Wu:08,Daghofer:08,Seo08,Chen08,Laad08,Kou08,Hackl08}.

\begin{figure}[htbp]
   \centering
\includegraphics[width=4.5in]{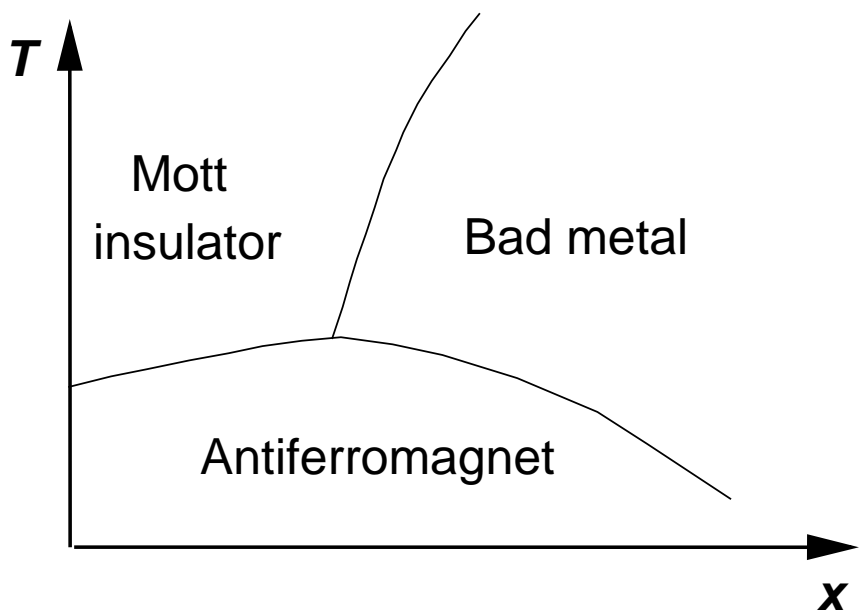}
  \caption{Schematic phase diagram of ${\rm V_2O_3}$ and ${\rm NiS_2}$.
$x$ could be external pressure or chemical doping (Cr or Ti 
doping for $V$; Se doping for S). The bad-metal regime is referred to
in Fig.~\ref{fig:correlation-placement} as the incipient-Mott regime.
The antiferromagnetic region may be further divided into insulating and 
metallic phases.
}
   \label{fig:mott-transition}
\end{figure}

Systems known \cite{McWhan_PRB73,Carter_PRL91,Matsuura_JPSJ00}
to be in this category
are ${\rm V_2O_3}$ and 
${\rm NiS_{2-x}Se_x}$,
whose phase diagram is shown schematically in
Fig.~\ref{fig:mott-transition}. 
At room temperature and ambient pressure, ${\rm V_2O_3}$
lies inside the paramagnetic metal part of its phase diagram
and displays all the features of a ``bad metal'' (see below).
Through the application of a negative chemical 
pressure (Cr doping for V), it can be explicitly tuned through
a localization transition into a Mott insulator state. 
Similarly, ${\rm NiS_{2-x}Se_x}$ can be tuned across 
the Mott transition as a function of the Se-doping for S.
The transition has been extensively studied in terms of 
the dynamical mean field theory (DMFT) \cite{Georges_rmp}. 

\section{Charge transport and dynamics}

A metal near a Mott transition will be a bad metal, which can 
be operationally defined in terms of a large electrical resistivity
at room temperature. The large resistivity corresponds to 
an effective mean free path $\ell$ that is on the order of the
average inter-electron spacing. It originates from the presence of 
incoherent electronic excitations, which represent the precursor 
to Mott localization. Note that our discussion of bad metal
shares some similarity with that of Ref.~\cite{EmeryKivelson},
although the issue of resistivity saturation at high temperatures
does not come into play in our consideration.

Experiments on the charge transport and charge dynamics 
indeed show that the iron pnictides are bad
metals. 
Consider first the electrical resistivity at room temperature;
it is on the order of 5 ${\rm m\Omega\cdot cm}$ for 
polycrystals \cite{Kamihara_FeAs}
and 0.5 ${\rm m\Omega\cdot cm}$ for single crystals \cite{Ni0806,Chen0806}.
These yield a 
$k_F\ell$ of order unity. 

Consider next the optical conductivity. The bad metal aspect is most
clearly seen in the reduced Drude weight.
Indeed, the Drude peak is hardly visible in 
polycrystal compounds \cite{Dong0803,Boris09}.
In single crystals, the Drude weight is relatively 
small \cite{Hu0806,Yang0807},
and the fitted plasma frequency is of the order of that extracted for
either ${\rm V_2O_3}$ or doped cuprates. 

The bad metal physics is also manifested in the higher-energy part 
of the optical spectra.
Since the development of the incoherent excitations comes at the expense 
of the coherent excitations (and vice versa),
spectral weight in the optical conductivity 
can redistribute between the relatively high energy range and the lowest 
energies.
Indeed, a relatively small change of temperature has been 
shown \cite{Hu0806,Yang0807,Boris09}
to induce a significant transfer of spectral weight between
the low-frequency Drude part and the part at high energies (0.6-1.5 eV).

All these features have strong similarities to what have been observed
in both ${\rm V_2O_3}$ and NiS$_{2-x}$Se$_x$.
At room temperature and ambient pressure, V$_2$O$_3$
lies inside the paramagnetic metal part of its phase diagram
({\it cf.}~Fig.~\ref{fig:mott-transition}),
and its electrical resistivity is of the order 
of 0.5 ${\rm m\Omega\cdot cm}$ \cite{McWhan_PRB73}.
In contrast, Cr is a good metal, with a room-temperature resistivity 
that is very small,
about 0.01 ${\rm m\Omega\cdot  cm}$ \cite{Yeh_nature02}.
When converted into a mean free path,
the value for ${\rm V_2O_3}$ is compatible to inter-particle spacing
while that for Cr is close to 100 times that of the inter-particle 
spacing.

Likewise, in ${\rm V_2O_3}$, the Drude weight is a relatively small part
of the total spectral weight and the fitted plasma frequency is 
about 1.3~eV \cite{Thomas_JLTP94}. 
As expected, the Drude peak is eliminated when ${\rm V_2O_3}$ is tuned into 
the Mott insulating state \cite{Thomas_PRL94}.
Finally, the temperature-induced 
transfer of spectral weight \cite{Rozenberg_PRL95,Baldassarre_PRB08}
between the Drude part and the higher-energy ($\sim$0.5--1.5~eV) part is 
well-established.
Similar evolution of the optical conductivity has also been
observed \cite{Perucchi08} 
in ${\rm NiS_{2-x}Se_x}$, across its Mott transition.

\section{Single-electron excitations}

\subsection{Coherent and incoherent excitations}

At the single-particle level, the coherent and incoherent excitations
will appear as a natural decomposition of the total density of states.
Photoemission measurements in principle provide a measurement of the
density of states, but the results for the iron pnictides 
are at this point still uncertain.

A recent angular-resolved photoemission (ARPES) 
measurement  on BaFe$_2$As$_2$~\cite{Ding_incoherent:0812}
has indicated incoherent excitations at very high energies (around 7 eV).
There are satellite features at about 2-3 eV range as well, and whether they
are entirely due to the interband contributions of non-interacting electrons,
or whether they involve the interaction-induced incoherent excitations 
as well, remains to be clarified.

Earlier ARPES measurements in LaOFeP, however, did not see any indication
of the incoherent features \cite{Lu_Nature08}. 
This could suggest that the correlation effects
are considerably weaker in LaOFeP.
Since the Fermi surface 
appearing in the ARPES measurement \cite{Lu_Nature08} 
is much larger than that appearing in 
a bulk de Haas - van Alphen (dHvA) experiment \cite{Coldea_PRL08}, 
an alternative possibility is that due to some surface contamination,
the states at the surface that are probed by ARPES are somewhat
overdoped \cite{Lu_Nature08}. In that case the incoherent features
will be significantly suppressed. This possibility could also reconcile 
the ARPES result with the optical conductivity measurements
in LaOFeP, which \cite{Qazilbash0808}
have shown features similar to those of the parent 
iron arsenides, namely reduced Drude weight and temperature-induced
transfer between low-energy and high energy spectral weights.

In ${\rm V_2O_3}$, again as a comparison,
both the coherent and incoherent features have been 
observed in photoemission experiments \cite{Mo_PRL03,Mo_PRL06}.

\subsection{Fermi surface}

The decomposition of the single-electron excitations into coherent
and incoherent ones can be realized by a decomposition of the $d$-electron
field operators, thus
$d =d^{coh} + d^{incoh}$, with $d^{coh}$ containing 
a fraction of spectral weight $w$, and $d^{incoh}$ containing 
$1-w$. The low-energy single-electron excitations are described 
in terms of $d^{coh}$, which can be normalized by $w$ through
$c=d^{coh}/\sqrt{w}$ (see below).

The coherent part of the single-electron Green's function is 
formally derived from the non-interacting Green's function
by a self-energy function. Various microscopic approaches
-- such as DMFT or slave-boson approaches -- are able to capture this
effect using a momentum ({\it i.e.} ${\bf k})-$independent self energy,
which 
leads to a Fermi surface that is identical to that 
of the non-interacting case. Conversely, the consistency of the 
measured Fermi surface with its non-interacting counterpart
(derived from LDA bandstructure calculations, for example)
only implies that the self energy is largely ${\bf k}-$independent,
and cannot by itself be used to infer 
the degree of electron correlations.

This last statement also applies to the antiferromagnetically ordered
state in the parent iron arsenides. In the incipient Mott picture,
even though the driving force
for the antiferromagnetism does not come from a Fermi-surface nesting,
the antiferromagnetic order parameter will on general grounds 
(and also through microscopic considerations, 
Sec.~\ref{Sec:Mag}) introduce
a staggered magnetic field, ${\bf h}_{\rm AF} ({\bf r})
= {\bf \Delta} e^{i {\bf Q} \cdot {\bf r}} $, which will be coupled
to the coherent itinerant electrons. This results in a coupling term,
\begin{eqnarray}
{\bf \Delta} \cdot \sum_{\bf k} 
M_{\alpha,\beta}
c_{{\bf k+Q} \alpha \sigma}^{\dagger}
{\btau}_{\sigma,\sigma'}
c_{{\bf k} \beta \sigma'}
\label{staggered_field}
\end{eqnarray}
where $M_{\alpha,\beta}$ specifies the details of the structure
in the orbital space ($\alpha,\beta,\ldots$),  
${\btau}$ are the Pauli matrices, and repeated symbols are summed.
This coupling has the same form as that in the case of an SDW state
and it leads to similar hybridized bands and the opening of the gap
centered around the ``hot spots'' on the Fermi surface (the Fermi
momenta ${\bf k_F}$ for which $\epsilon_{\bf k_F} = 
\epsilon_{{\bf k_F}+{\bf Q}}$).
Precisely this type of band reconstruction 
has recently been reported in an 
ARPES experiment \cite{Hsieh_0812}, which is also largely consistent 
with the dHvA measurement \cite{Sebastian_JPCM08}. Our discussion
shows that the reconstruction cannot be used to infer the strength 
of correlation effects.

We comment on the effective mass of the low-energy excitations;
it will be enhanced compared to the LDA bandstructure value.
But the effective mass extracted in many measurements -- including
thermodynamics and single-particle dispersion -- will be smaller 
than $1/w$ because of the non-zero exchange coupling (see the next
section). The mass enhancement (measured with respect to the band 
mass at the Fermi energy) estimated from dHvA \cite{Sebastian_JPCM08}
is about 2, and that 
from ARPES ranges from 2-4 and can be orbitally 
selective \cite{Ding_incoherent:0812}.
Separately, ARPES measurements have also been used to estimate the 
band narrowing [defined by the ratio of the LDA-calculated value 
to the measured value, for the separation
between the bottom of the bands containing the electron pockets near
the $M$ point $\equiv (\pi,0)$ and the top of the bands containing the hole
pockets near $\Gamma \equiv (0,0)$]
to be about 5 \cite{Yang0806,Ding_incoherent:0812}.

\section{Magnetic order and dynamics}
\label{Sec:Mag}

We now derive the low-energy effective Hamiltonian.
We can separately consider the on-site Coulomb interactions,
including the Hund's coupling, and the kinetic energy.
The interactions at each site can be diagonalized into local
configurations, which naturally divide into low energy and high
energy sectors. A typical kinetic term of the Hamiltonian has the
form of $H_t=\sum t_{ij}^{\alpha\beta}
d_{i \alpha \sigma}^{\dagger}
d_{j \beta \sigma}$. With the decomposition 
$d =d^{coh} + d^{incoh}$, $H_t$
can be rewritten as the sum of 
three terms:
\begin{eqnarray}
H_{t1} &=& \sum t_{ij}^{\alpha\beta} d_{i \alpha \sigma}^{\dagger coh}
d_{j \beta \sigma}^{coh} \nonumber \\
H_{t2} &=& \sum t_{ij}^{\alpha\beta} d_{i \alpha \sigma}^{\dagger incoh}
d_{j \beta \sigma}^{incoh} \nonumber \\
H_{t3} &=& \sum t_{ij}^{\alpha\beta} \left ( d_{i \alpha \sigma}^{\dagger coh}
d_{j \beta \sigma}^{incoh} + H.c. \right )
\label{H_t_123}
\end{eqnarray}

We can now discuss the form of the low-energy effective Hamiltonian.
$d_{i \beta \sigma}^{coh}$ 
describes low-energy electronic degrees of freedom,
operating entirely
within the low-energy
sector of the many-body spectra,
so $H_{t1}$ is part of the low-energy model.
$d_{i \beta \sigma}^{incoh}$, on the other hand,
represents high-energy electronic degrees of freedom,
connecting 
the low energy and high energy
sectors of the many-body spectra,
so we need to project out the high energy
sectors in both $H_{t2}$ and $H_{t3}$.

To do so, we will adopt the projection procedure of Ref.~\cite{Moeller:95}.
Here, $d_{{\bf k}\alpha\sigma}^{coh}$ appears as the $d$-electron operator
projected to the coherent part of the electronic states near
the Fermi energy. (For readers who are inclined to think 
in terms of the slave-boson-type representation of an electron
operator, $d_{{\bf k}\alpha\sigma}^{coh}$ corresponds to the part
with the slave-boson replaced by its condensate amplitude.)
Unlike the full $d$-electron operator, $d_{{\bf k} \alpha \sigma}^{coh}$
does not satisfy the fermion anticommutation relation:
its spectral density integrated over frequency defines $w$.
It will be convenient to introduce
$c_{{\bf k} \alpha \sigma}=(1/\sqrt{w})d_{{\bf k} \alpha \sigma}^{coh}$,
which has a total spectral weight normalized to $1$ and which 
satisfies $\{c_{{\bf k} \alpha \sigma}, 
c^{\dagger}_{{\bf k}\alpha \sigma}\}=1$.
We note that $<n|d^{coh}|m>$ and $<m|d^{\dagger incoh}|n>$ cannot be
simultaneously nonzero for any pair of many-body states $|n>$ and $|m>$;
in other words, $<n|d^{coh}|m>$ is nonzero only when both 
$|n>$ and $|m>$ belong to the low-energy sector, while 
$<m|d^{\dagger incoh}|n>$ is nonzero only when one of $|n>$ and $|m>$
belongs to the low-energy sector while the other belongs to 
the high-energy sector. Correspondingly, the single-electron
Green's function separates into a coherent part and an incoherent one,
and so does the single-electron spectral density.
We also note that, in the $w \rightarrow 0$ limit, we reach the Mott
transition point where there are only incoherent excitations. 
For the spectral density of the incoherent excitations at the
Mott transition, we will assume  \cite{Moeller:95} (based on the
DMFT results \cite{Georges_rmp}) that it is gapped.

To the leading non-vanishing order in $w$,
$H_{t2}$ is then 
adiabatically connected to the form of the kinetic
term deep in the Mott insulating state. The notion of superexchange
applies: integrating out the high energy sector of the many-body spectra
here will result in a multi-band matrix Heisenberg 
type of Hamiltonian, $H_J$. This was done in Ref.~\cite{Si_prl08}.
A key feature that differentiates the iron pnictides from other 
metals lying at the boundary of localization is that $H_J$ must
incorporate exchange interactions beyond the nearest-neighbors (n.n.)
within the square lattice of the Fe atoms. Indeed, because
each As atom has an equal distance from each of the four Fe atoms
in a square plaquette, the next-nearest-neighbor 
interaction, $J_2$, is important in addition to the 
nearest-neighbor interaction, $J_1$, This leads to
\begin{eqnarray}
H_J  
=
\sum_{\langle ij\rangle} J_1^{\alpha\beta}
{\bf s}_{i,\alpha} \cdot {\bf s}_{j,\beta}
+
\sum_{\langle\langle ij\rangle\rangle} J_2^{\alpha\beta}
{\bf s}_{i,\alpha} \cdot {\bf s}_{j,\beta}
+
\sum_{i,\alpha \ne \beta} 
J_H ^{\alpha\beta}
{\bf s}_{i,\alpha} \cdot {\bf s}_{i,\beta} .
\label{H_J}
\end{eqnarray}
where $\langle ij\rangle$ and 
$\langle\langle ij\rangle\rangle$ label 
nearest-neighbor (n.n.) and next-nearest-neighbor (n.n.n.)
Fe sites on its square lattice.
Here, the ${\bf s}$ operators are the spins associated with
the incoherent part of the electronic excitation; they represent
the total electron spin when $w$=0.
For convenience, we have also grouped $J_H$,
the on-site Hund's coupling, in $H_J$.
Because of the multiple orbitals, both $J_1$ and $J_2$ 
are matrices. 
On general grounds \cite{Si_prl08},
the largest eigenvalue of $J_2$ is expected to 
be somewhat larger than that of $J_1/2$. 
Both of these largest eigenvalues are antiferromagnetic in nature.

$H_{t1}$ and $H_{t3}$ can be recast in a convenient form 
by using 
$c_{{\bf k} \alpha \sigma}$ introduced earlier.
After Fourier transform,
$H_{t1}$ then becomes
\begin{eqnarray}
H_c = w \sum_{{\bf k},\alpha,\sigma} E_{{\bf k}\alpha\sigma}
c_{{\bf k} \alpha \sigma}^{\dagger}
c_{{\bf k} \alpha \sigma}
\label{Hc}
\end{eqnarray}
Finally, projecting the high energy sector of the many-body spectra
of $H_{t3}$ leads to the following
coupling between the local moments and the 
coherent itinerant carriers:
\begin{eqnarray}
H_m = w \sum_{{\bf k} {\bf q} \alpha\beta\gamma} 
G_{{\bf k},{\bf q}\alpha\beta\gamma}~ 
c_{{\bf k}+{\bf q}\alpha\sigma}^{\dagger} 
\frac{\btau_{\sigma\sigma^{\prime}}}{2} 
c_{{\bf k}\beta\sigma^{\prime}} \cdot {\bf s}_{\mathbf{q}\gamma} \;,
\label{Hcouple}
\end{eqnarray}
where $\btau$ are the Pauli matrices,
and $G$ is generically antiferromagnetic.
The final form for the low-energy effective Hamiltonian is
\begin{eqnarray}
H_{eff} = H_J + H_c + H_m .
\label{Heff}
\end{eqnarray}

The $J_1$-$J_2$ exchange interactions 
in the mentioned range lead to 
a ${\bf Q} = (\pi,0)$
collinear antiferromagnet, which is precisely what 
has been observed in the parent iron arsenides 
\cite{Cruz:08,Qiu:08,Chen_Y:08,Kimber:08,Zhao_PrFeAsO,Zhao:08,Huang:08,Jesche:08}.
Moreover, they cause magnetic frustration which will in turn 
reduce the ordered moment. (Numerical results on the square-lattice
$J_1-J_2$ model have shown that the frustration-induced 
suppression of the ordered moment can be substantial \cite{Darradi08}.)
This is also consistent with 
neutron-scattering observation that the ordered moment in the 
parent iron arsenides can be as small 
as $0.3\mu_B/$Fe \cite{Cruz:08,Qiu:08,Chen_Y:08,Kimber:08,Zhao_PrFeAsO,Zhao:08,Huang:08,Jesche:08}, 
as opposed to the order of $2\mu_B$/Fe expected from either 
ionic considerations or LDA-based 
{\it ab initio} calculations.

Another appealing feature that follows from the $J_1$-$J_2$ model 
is that it leads to an Ising order \cite{Fang:08,Xu:08,Chandra}.
The collinear order can be 
either ${\bf Q} = (\pi,0)$ or ${\bf Q'} = (0,\pi)$, with the
corresponding magnetic order parameters ${\bf m}$ or ${\bf m'}$.
The composite scalar, ${\bf m} \cdot {\bf m'}$, is an Ising order 
parameter. It is very natural to interpret the structural transition,
observed in the parent iron arsenides as originating from a coupling
of certain structural degrees of freedom with the Ising order parameter.

The effective Hamiltonian, Eqs.~(\ref{H_J},\ref{Hc},\ref{Hcouple},\ref{Heff}),
provides the basis to specify the notion of $J_1$-$J_2$ magnetic
frustration even though the system is metallic.
Note that the $J_1$-$J_2$ couplings have also been 
studied \cite{Yildirim,Ma,Yin} using {\it ab initio} calculations.
Experimentally, spin waves in the magnetically
ordered arsenides are being measured 
using the inelastic neutron scattering technique
\cite{Zhao0808,Ewings0808,McQueeney0809}.
While the results are still preliminary, 
they have allowed an interpretation based on the $J_1$-$J_2$ model,
yielding exchange interactions on the order of 40 meV.

Microscopic studies of the effective Hamiltonian,
Eqs.~(\ref{H_J},\ref{Hc},\ref{Hcouple},\ref{Heff}), remain
to be carried out. 
This Hamiltonian has the form of a Kondo lattice, in which 
the Kondo-like coupling of Eq.~(\ref{Hcouple})
is in competition with antiferromagnetic exchange interactions among 
the ``local moments'' of the incoherent part of the spectrum,
Eq.~(\ref{H_J}). A microscopic analysis should 
determine $w$ as a function of the bare interactions,
and constrain the spectral weight at the Fermi level 
to be the coherent itinerant carriers; the latter will ensure
that the Fermi volume satisfies the Luttinger theorem.

For the  purpose of analyzing the magnetic phase diagram of 
the parent compounds, it would then suffice to use $w$ 
-- as opposed to the bare interactions -- as the non-thermal
control parameter and to proceed with an effective field
theory \cite{Dai_08}. A small $w$ does not change the magnetic 
structure resulting from Eq.~(\ref{H_J}). Increasing $w$ leads
to a magnetic quantum critical point (QCP), to which we now turn.

\section{Quantum criticality}

For $w=0$, the magnetic order is captured by the following 
Ginzburg-Landau action:
\begin{eqnarray}
{\cal S}
\! & = &\!\! 
\int d {\bf q}
\! \! \int 
\!\! d \omega
\left[ 
(r +\omega^2 + c {\bf q}^2 )
((\, {\bf m} \,)^2 + (\, {\bf m'} \,)^2)
+  v (q_x^2-q_y^2) {\bf m}\cdot{\bf m'}
 \right]
\nonumber\\
&+&
\int \!\! 
\left [ ~\!\!\, u \, (\, {\bf m} \,)^4 
+ \!\!\, u \, (\, {\bf m'} \,)^4 
+ \!\!\, \tilde{u} \, ({\bf m} \cdot {\bf m}')^2
+ \!\!\, u' \, ({\bf m})^2({\bf m'})^2
\!\! 
~\right ]
+ \ldots ,
\label{S-effective-0}
\end{eqnarray}
in which $r<0$ places the system on the magnetically ordered side
of the phase diagram. The coupling terms of this 
quantum sigma model have a similar form as those of 
their classical counterparts \cite{Chandra}.

Non-zero $w$ couples the local moments to the coherent itinerant
carriers, as specified by Eq.~(\ref{Hcouple}). Its primary effect on the 
Ginzburg-Landau action is to modify the quadratic coefficient,
by introducing a positive shift to $r$ and a damping term.
To the leading non-vanishing order in $w$, the addition to $\cal S$ is
\begin{eqnarray}
\Delta {\cal S}\! & = &\!\! 
\int d {\bf q}
\! \! \int \!\! d \omega
( w A_{\bf Q} + \gamma |\omega| )
\left[ (\, {\bf m} \,)^2 + (\, {\bf m'} \,)^2
\right ]
\label{S-effective-w}
\end{eqnarray}
The linear-in-frequency damping term, $\gamma |\omega|$,
arises because ${\bf Q} = (\pi,0)$ connects the electron and hole
pockets of the Fermi surface.
This form is valid for $|\omega |$ up to
$w W$, where $W$ is the effective bandwidth 
of the non-interacting electrons; it is useful to emphasize that 
the leading term in $\gamma$ is of the order $w^0$,
{\it i.e.} $O(1)$ instead of $O(w)$.
Note that, our purpose of considering the Landau damping 
is primarily to discuss quantum criticality (see below),
and we have correspondingly constructed the damping term
appropriate for the magnetically disordered case.

The $w A_{\bf Q}$ term in Eq.~(\ref{S-effective-w}),
where $A_{\bf Q}$ is a positive constant,
is a shift to the mass term of the Ginzburg-Landau theory.
Increasing $w$ makes $r(w) = r +wA_{\bf Q}$ less negative,
weakening the magnetic order.
This provides a natural account of the neutron-scattering observation 
\cite{Cruz:08,Qiu:08,Chen_Y:08,Kimber:08,Zhao_PrFeAsO,Zhao:08,Huang:08,Jesche:08}
that the ordered staggered moment considerably varies across
the parent arsenides, ranging from about 0.3$\mu_B$/Fe to about 0.9$\mu_B$/Fe.
This effect is in addition to 
the variation of the ordered moment due to a change of the $J_2/J_1$
ratio among the parent arsenides.

Moreover, it is expected that, $r(w_c)=0$ at a non-zero $w_c$, 
which corresponds to an antiferromagnetic quantum critical point (QCP).
The linear-in-$\omega$ damping implies that the dynamic exponent 
$z=2$. One proposal to explore such a magnetic QCP is to use 
P-substitution for As in the parent arsenides \cite{Dai_08}.
Experimentally, independent considerations have led to 
work on the CeOFeAs$_{\rm 1-\delta}$P$_{\rm \delta}$ 
series \cite{Geibel}. It is hoped that it will soon become experimentally
feasible to ascertain the degree to which magnetic quantum criticality
is realized in the parent iron arsenides.

We note in passing on a particularly intriguing aspect of quantum
criticality. Our discussion of the magnetic QCP has focused on 
the type that conforms to the standard theory based on
order-parameter fluctuations. It will be interesting to see whether
more unconventional type of quantum criticality, such as those
involving a sharp jump of the Fermi surface \cite{Gegenwart08},
can come into play in this context.

A magnetic QCP may in principle also occur as a function of 
carrier doping, which also enhances quantum fluctuations.
Carrier doping, however, introduces several features that
complicate quantum criticality.
First, how carrier doping modifies the Fermi surfaces
remains uncertain. In particular,
if the magnetic ordering vector ${\bf Q}$
no longer connects the electron and hole pockets,
the linear-in-$\omega$
form of the damping term in Eq.~(\ref{S-effective-w})
may be absent, and the magnetic QCP will be pre-emptied by 
a QCP associated with the Ising-transition-induced structural
distortion \cite{Xu:08,Xu0812}.

Second, the very fact that carrier doping leads to optimal 
superconductivity impedes the observation of quantum criticality.
Indeed, a variety of
behaviors for the zero-temperature phase change from antiferromagnetism 
to superconductivity \cite{Zhao:08,Drew0807,Luetkens0806,Chen0807},
have been observed. 
The complex interplay between magnetic quantum
transition and superconductivity is well known, for example 
in heavy
fermion systems, where there are concrete cases -- particularly
the Rh-based 115 compound -- in which a well-defined QCP gets 
exposed only when superconductivity is 
suppressed \cite{Park06, Knebel06,Knebel08,Gegenwart08}.
Unfortunately, the standard means of suppressing superconductivity 
-- applying a magnetic field -- is not very easy to implement 
in the iron pnictides, because the upper critical field $H_{c2}$
appears to be rather high.
It is for this reason that focusing on the parent compounds may be more
advantageous, where indications are that superconductivity is not as strong
as in the doped cases.

Still, there is one particularly appealing feature with the alkali-metal 
doped 122 system. In ${\rm K_{1-x}Sr_xFe_2As_2}$, the doping $x$ can 
cover the entire range from $0$ to $1$ \cite{Sasmal08}. 
This opens up a large range of the control parameter 
space, which is advantageous
for identifying the signatures of quantum criticality. Indeed, 
recent measurements of the electrical and thermoelectric transport properties
in this series have provided some support for quantum critical fluctuations 
in the vicinity of the optimal doping \cite{Gooch0812}.

%
%

\section{Multiband matrix \textbf{\textit{t-J$_{\bf 1}$-J$_{\bf 2}$}} model}
From the microscopic point of view, the notion that the parent iron
pnictides are in proximity to a Mott insulator provides the basis 
to consider the doped iron pnictides in terms of a strong coupling
approach. The multiband matrix $t$-$J_1$-$J_2$ model has the 
form \cite{Si_prl08}:
\[
H_{t-J_1-J_2}= H_t + H_{J_1-J_2}, \nonumber
\]
where
\begin{eqnarray}
H_t&=& \sum_{ij} t_{ij}^{\alpha\beta}
{\tilde{c}}_{i,\alpha}^{\dag} {\tilde{c}}_{j,\beta} \cr
H_{J_1-J_2}&=&\sum_{\langle ij \rangle}  
J_1^{\alpha\beta}
{\bf s}_{i,\alpha} \cdot {\bf s}_{j,\beta}
+ \sum_{\langle\langle
ij\rangle\rangle} 
J_2^{\alpha\beta}
{\bf s}_{i,\alpha} \cdot {\bf s}_{j,\beta}
+ 
\sum_{i,\alpha \ne \beta} 
 J_H ^{\alpha\beta}
{\bf s}_{i,\alpha} \cdot {\bf s}_{i,\beta} .
\label{H_t_J1_J2}
\end{eqnarray}
Here the multiple bands are labeled by the orbital indices $\alpha,\beta$ 
as before
and ${\tilde{c}}_{\alpha,i}$ describe the constrained fermions at the 
site $i$. The n.n. and n.n.n.
hybridization matrices
are $t_{n.n}^{\alpha\beta}=t_1^{\alpha\beta}$
and $t_{n.n.n}=t_2^{\alpha\beta}$.
As noted,
the n.n. ($\langle ij \rangle $) and n.n.n. ($\langle
\langle ij \rangle\rangle $)
superexchange matrices are 
$J_{n.n}^{\alpha\beta}=J_1^{\alpha\beta}$ and 
$J_{n.n.n.}^{\alpha\beta}
=J_2^{\alpha\beta}$.

This model can be used to study a number of issues that bear on the
correlated-electron
physics of the carrier-doped iron pnictides.
An intriguing possibility is that a carrier-doped state from
a parent system with a non-zero but small $w$ behaves similarly
to that from a parent system with $w=0$.

\section{Summary}

Our discussions above lead to the picture that the iron pnictides have 
intermediately-strong electron correlations, and are located at the 
boundary between localization and itinerancy. The proximity to the Mott
insulating state separates the single-electron excitations into coherent 
and incoherent ones, which in turn provide the basis to discuss magnetism
and magnetic fluctuations in terms of a combination of local moments
and itinerant carriers. Tuning the degree of itinerancy provides a means
to vary the strength of magnetic ordering, and leads to a magnetic 
quantum critical point. Finally, this picture also provides the basis
to discuss the carrier-doped iron pnictides in terms of a multiband matrix
$t$-$J_1$-$J_2$ model.

This work has been supported in part by
the NSF Grant No. DMR-0706625 and the Robert
A. Welch Foundation (Q.S.),
the NSF of China and PCSIRT (IRT-0754) of Education
Ministry of China (J.D.), and 
the U.S. Department of Energy (J.-X.Z.).

\end{document}